\documentstyle[12pt]{article}
\advance\textheight by \topskip
\begin{document}

\begin{flushright}
CERN-TH-97-17 \\
DAMTP R-97-09\\
SU-ITP-97-05\\
hep-th/9702103\\
February 1997\\
\end{flushright}
\vspace{.5cm}

\begin{center}
\baselineskip=16pt

{\Large\bf  BLACK HOLES AND   CRITICAL POINTS  \\

\

 IN MODULI SPACE}

\vskip 1 cm

{\bf  Sergio Ferrara,$^a$~
Gary W. Gibbons,$^b$~ and \,Renata Kallosh$^c$
}\footnote{E-mail:
FERRARAS@vxcern.cern.ch,~~G.W.Gibbons@damtp.cam.ac.uk,\\
\indent ~~kallosh@physics.stanford.edu}

\vskip 1cm

$^a${\em Theory Division, CERN, 1211 Geneva 23, Switzerland}\\
$^b${\em DAMTP,  Cambridge University, Silver Street, Cambridge CB3 9EW,
United Kingdom}\\
$^c${\em Physics Department, Stanford
University, Stanford, CA 94305-4060, USA}

\end{center}

\vskip 1 cm
\centerline{\bf ABSTRACT}
\vspace{-0.3cm}
\begin{quote}
We study  the  stabilization of scalars near a
supersymmetric black hole horizon using the  equation of motion
of a particle
moving in a potential and background metric. When the relevant 4-dimensional
theory is described by
 special geometry, the generic properties of the critical points of this
potential can be studied.  We find that the extremal value  of  the
central charge  provides the minimal value of the BPS mass and of the potential
under the condition that  the moduli space metric is positive
at the critical point. This is a property of a regular special geometry. We
also study the critical points in  all
N$\geq $2 supersymmetric theories.  We relate these ideas to the
Weinhold and Ruppeiner metrics introduced in  the geometric approach to
thermodynamics and used for study of critical phenomena.

\end{quote}
\normalsize
\newpage


\section{Introduction}
In this paper we intend to tie together some recent (and not so recent) work on
4-dimensional black holes in N=2 ungauged supergravity theories
\cite{gary}-\cite{KLMS} . These theories have {\it two types of geometries}:
space-time geometry and moduli space geometry, so called special
geometry \cite{special}-\cite{Cer1}  in the space of the scalar fields of the
theory. Various properties of space-time singularities including black holes
have been studied for a long time. Much less is known about the singularities
of the moduli space. In view of the recent understanding that there is a web of
connections between different versions of string theories and supergravities
one  can view the study of these two type of geometries as a useful tool for
clarifying such connections. An interesting example of an interplay between the
two type of singularities is provided by the massless black holes \cite{strom}.
Such
solutions have been constructed and studied before \cite{massless,KL} in the
heterotic string theory and have found to have naked singularities \cite{KL}.
More recently these solutions have been reexamined with account taken of the
first loop corrections of  the heterotic string \cite{klaus}. These corrections
modify the prepotential of N=2 supergravity model from $STU +  a U^3$.
The  massless black holes with some  charges negative and some charges positive
have the following time components of the metric
\begin{equation}
 g_{tt}^2 =  4 (h_0 + {q_0 \over r}) \left(
 (h^1 - {p^1 \over r})  (h^2 - {p^2 \over r})
 (h^3 + {p^3 \over r}) + a (h^3 + {p^3 \over r})^3 \right) \ .
\end{equation}
The classical solution ($a=0$) has some naked
 singularity  which makes the black hole repulsive to all matter
\cite{KL}. In the internal space this singularity corresponds to a
vanishing cycle.  If one includes the quantum
corrections one may remove the naked singularities from the space-time by a
proper choice of the
parameters, still keeping the mass vanishing. The quantum correction
proportional to $a$  seem to act  as a
regulator which removes some of the singularities of the space-time metric.
However by looking into the  scalar metric in the moduli space  $g_{k\bar k}  =
\partial_k \partial _{\bar k} K(z, \bar z)$ one can find that some components
of the scalar metric given by the second derivative of the Kahler potential
become negative, at least for  $a={1\over 3}$ which is the actual number coming
from the first loop calculation in  heterotic string theory. This signals that
the
moduli space geometry becomes singular as the price for having the space-time
geometry  singularity free. Note that the change in  sign of the
scalar metric without a change of the signature of the space-time means that
the Lagrangian has the wrong sign for the kinetic term for scalars. Moreover
the
fact that the sign of the scalar metric changes from the positive to the
negative one means that somewhere in the space-time the metric vanishes and the
special geometry is singular.

In general it may be useful to study the properties of these two geometries
together. The purpose of this paper is to set up the relevant connections
between space-time and special geometry.
In particular this will allow us to establish the properties of the critical
points of the central charge in supersymmetric theories. We will find out when
it is a minimum, when it is a maximum and whether  one should expect the
uniqueness or non-uniqueness of the critical points. Without the use of the
special geometry in the moduli space one can only try to address these issues
on a case by case basis. However special geometry will allow us to have a clear
answer to all these problems.

The generalization of this study to the higher supersymmetries $N>2$ is also
possible using the recently developed new formulation of extended
supergravities with build-in symplectic structure \cite{AAF}.
The basic difference between N=2 and $N>2$ is in the structure of the moduli
space: for N=2 the scalar manifold in not necessarily a coset manifold whereas
for $N>2$ it is a coset manifold.

Upon identifying the second derivative of the BPS mass and of the black hole
potential
with the metric on the moduli space of N=2 theory we have realized the deep
connection of this construction to the geometric approach to thermodynamics
where the
Weinhold and Ruppeiner metrics were used for the study of critical phenomena
\cite{rup}.

In Sec. 2 we mainly review some work on black holes and one-dimensional
geodesic motion which was introduced earlier \cite{gary} and adapt this work
with the
purpose of relating it to the special geometry. In Sec. 3 we focus on extreme
and double-extreme configurations and find  that equations of motion of
a particle in a potential together with regularity requirement are sufficient
to explain the attractor mechanism for the scalars near the horizon which was
understood before \cite{FKS,FK}  with the help of supersymmetry. Sec. 4
contains new results: we identify the potential of N=2 theory with the first
symplectic
invariant of special geometry which is homogeneous and of degree 2 in electric
and magnetic charges. The critical points of the central charge are found to be
also the critical points of the potential.
We find the correlation between the sign of the second derivative of the
potential at the critical point and the sign of the scalar metric at a given
critical point. This relates the singularities of the moduli space to the
critical behavior of the potentials defining the motion in space-time. In Sec.
5 we study the critical points of the BPS mass and of the black holes potential
in arbitrary extended supergravity with or without matter and calculate the
second derivatives at the critical points. Sec. 6 is devoted to connection to
the third geometry used in
geometric approach to thermodynamics.

\section { Geodesic Action with a Constraint}
The class of theories we wish to consider has Lagrangian

\begin{equation}  -{R\over 2} +\frac{1}{2} G_{ab} \partial_ \mu \phi ^a
\partial_\nu \phi ^b
g^{\mu\nu}
 -\frac{1}{4}
 \mu_{\Lambda \Sigma} {\cal F}^{\Lambda}_{\mu \nu}  {\cal
F}^{\Sigma}_{\lambda \rho}  g^{\mu \lambda} g^{\nu \rho}
- \frac{1}{4} \nu_{\Lambda \Sigma} {\cal F}^{\Lambda}_{\mu \nu}{}^*{\cal
F}^{\Sigma}_{\lambda \rho} g^{\mu \lambda} g^{\nu \rho}
 \ .
\label{scalaraction}\end{equation}
We restrict attention to static solutions and make the ansatz for the metric in
the form
\begin{equation}
ds^2 = e^{2U} dt^2 - e^{-2U} \gamma_{mn} dx^m dx^n \ .
\end{equation}
The effective 3-dimensional Lagrangian from which the field equations can be
derived takes the form
\begin{equation}
{1\over 2}R[\gamma_{mn}] - {1\over 2}  \gamma^{mn} \partial_m \hat \phi ^a
\partial_n \hat \phi
^b \hat G_{ab} \ ,
\end{equation}
where the ``hatted " scalar fields include in addition to the scalar fields
$\phi^a$ of the 4-dimensional theories  also the function $U$ defining the
metric as well as the electrostatic $\psi^A$ and magnetic static  $\chi_A$
potentials
\begin{equation}
 \hat \phi ^a = ( U, \phi ^a,    \psi^\Lambda,  \chi_\Lambda )  \ .
\end{equation}
The metric $\hat G$ of the enlarged scalar manifold is independent of
$\psi^\Lambda$
and   $\chi_\Lambda$ as required by gauge independence.
Now consider spherically symmetric solutions. We specify the ansatz
\begin{eqnarray}
\gamma_{mn} dx^m dx^n  =  {c^4 d\tau^2 \over \sinh^4 c\tau}
+ {c^2
\over \sinh^2 c\tau}
(d \theta ^2 +  \sin^2 \theta d \varphi ^2) \ ,
\end{eqnarray}
and discover that the effective 1-dimensional Lagrangian from which the radial
equations may be derived is a pure geodesic action:
\begin{equation}
\hat G_{ab} {d \hat \phi ^a  \over d\tau } {d \hat \phi ^b  \over d\tau }  \ ,
\end{equation}
together with the constraint that
\begin{equation}
\hat G_{ab} {d \hat \phi ^a  \over d\tau } {d \hat \phi ^b  \over d\tau } = c^2
\ .
\end{equation}
Because $\hat G$  is independent of $\psi^\Lambda$ and   $\chi_\Lambda$ due to
gauge
invariance, we have constants of motion:
\begin{eqnarray}
p^\Lambda  &=&\hat G^{\Lambda \Sigma} {d \hat \chi  _\Sigma  \over d\tau }\ ,
\nonumber\\
 \nonumber\\
q_\Lambda &=&\hat G_{\Lambda \Sigma} {d \hat \psi^\Sigma  \over d\tau }\ .
\end{eqnarray}
 We may now replace the pure geodesic action by the
\begin{equation}
\left ({d U \over d\tau}\right )^2 + G_{ab}  {d\phi^a \over d\tau }
{d\phi^b
\over d\tau } +e^{2U}
V(\phi, (p,q))
 \end{equation}
and the constraint by

\begin{equation}
\left ({d U \over d\tau}\right )^2 + G_{ab}  {d\phi^a \over d\tau }
{d\phi^b
\over d\tau } - e^{2U} V(\phi, (p,q))  = c^2\ .
 \label{constr}\end{equation}
Here $V(\phi, (p,q))$ is a particular potential function constructed from the
scalar dependent positive definite couplings $\mu_{\Lambda \Sigma}$ and
$\nu_{\Lambda \Sigma}$ of vector fields and $c^2 = 2ST$, where $S$ is the
entropy and $T$
is the temperature of the black hole \cite{GKK}.
Specifically
\begin{equation}
V= {1\over 2}(p,q)   {\cal M}  \left (\matrix{
p\cr
q\cr
}\right ),
\label{pot}\end{equation}
where
\begin{equation}
 {\cal M} = \left |\matrix{
\mu+ \nu \mu^{-1} \nu  & \nu \mu^{-1}  \cr
\mu^{-1} \nu &  \mu^{-1} \cr
}\right |.
\label{Matrix}\end{equation}
 It is clear that the properties of the black holes in theories of this type
are governed entirely by the metric $G_{ab}$ on the scalars and the potential
function $ V(\phi, (p,q)) $.

\section{Extreme and Double-Extreme Holes}

We begin by considering extreme holes when $c^2 = 2ST=0$ and $\gamma_{mn} dx^m
dx^n $
is given by
\begin{eqnarray}
\gamma_{mn} dx^m dx^n  =  {d\tau^2 \over \tau^4}
+ {1
\over \tau^2}
(d \theta ^2 +  \sin^2 \theta d \varphi ^2) \ .
\end{eqnarray}
The geometry becomes
\begin{equation}
ds^2 = - e^{2U} dt^2 + e^{-2U} \left[{d\tau^2 \over \tau^4}
+ {1
\over \tau^2}
(d \theta ^2 +  \sin^2 \theta d \varphi ^2)\right]  \ .
\end{equation}
Evidently to obtain finite area solution we must have that
\begin{equation}
e^{-2U}\rightarrow \left( {A\over 4\pi}\right) \tau^2  \qquad {\rm as} \quad
\tau  \rightarrow - \infty \ .
\end{equation}
 We also require that this expression for our solution is not
infinite near the horizon,
\begin{equation}
G_{ab} {d \phi ^a \over  d\tau} {d  \phi ^b \over d\tau} e^{2U}  \tau^4 <
\infty \ .
\end{equation}
This leads to
 \begin{equation}
G_{ab} {d \phi ^a \over d\tau} {d  \phi ^b \over d\tau} \left( {4\pi \over A}
\right)   \tau^2 \rightarrow  X^2  \qquad {\rm as} \quad \tau \rightarrow -
\infty\ .
\end{equation}
Now we can substitute this into the constraint and we get
  \begin{equation}
{1\over \tau^2} +  \left( { X^2 A\over \tau^2  4\pi}\right) - {4\pi \over A}
{V(p,q,\phi_h)\over \tau^2}=0 \ .
\end{equation}
This leads to
\begin{equation}
A \leq 4\pi V(p,q,\phi_h)
\end{equation}

The near horizon geometry becomes equal to
\begin{equation}
ds^2 = -  {4\pi \over A\tau^2}  dt^2 +   \left( {A\over 4\pi}\right)
\left[{d\tau^2 \over \tau^2}
+ (d \theta ^2 +  \sin^2 \theta d \varphi ^2)\right]  \ .
\end{equation}
It is useful to change the variables as follows:
\begin{equation}
\rho= -{1\over \tau}\ , \qquad \omega = \log \rho
\end{equation}
and bring the near horizon geometry to the form of the product space
$AdS_2\times S^2$:
\begin{equation}
ds^2 = -  {4\pi \over A}  e^{2\omega} dt^2 +   \left( {A\over 4\pi}\right)
d\omega^2
+ \left( {A\over 4\pi}\right)  (d \theta ^2 +  \sin^2 \theta d \varphi ^2)  \ .
\end{equation}

In these coordinates the value of the derivatives of the moduli as the function
of $\omega$ enters in the term $G_{ab} \partial_ \mu \phi ^a  \partial_\nu \phi
^b g^{\mu\nu}
$ as follows
 \begin{equation}
G_{ab} {d \phi ^a \over d\omega } {d  \phi ^b \over d\omega } \left( {4\pi
\over A} \right)   \rightarrow   X^2  \qquad {\rm as} \quad \omega \rightarrow
 \infty\ .
\end{equation}
Now it is obvious that only $X^2=0$ is consistent with the requirement that the
moduli do not blow up near the horizon. Indeed, if
\begin{equation}
{d \phi ^a \over d\omega } = {\rm const} \qquad {\rm as} \quad \omega
\rightarrow  \infty
\end{equation}
and the moduli near the horizon have to be linear in $\omega$ they would not be
finite near the horizon. Thus we have proved that for $c=0$ extremal black
holes from the single requirement that the geometry as well as moduli are
regular near the horizon in the suitably chosen coordinates it follows that the
area of the horizon equals the value of the potential near horizon.
\begin{equation}
{A\over 4\pi} =  V(p,q,\phi_h) \ .
\label{equal}\end{equation}
This property near the horizon applies both to extremal and double-extremal
black holes. Double-extremal black holes \cite{KSW}  have the constant moduli,
so for them the term $G_{ab} \partial_ \mu \phi ^a  \partial_\nu \phi ^b
g^{\mu\nu}
$ vanishes everywhere and the equality (\ref{equal}) between the area of the
horizon and the value of the potential near the horizon follows from the
constraint equation (\ref{constr}) immediately. Thus we have also shown that
the area of the horizon of extreme black holes coincides with the area of the
horizon of the double-extreme black holes with the same values of charges and
is given by the value of the potential
\begin{equation}
A_{\rm extr} =A_{\rm double-extr}  =  4\pi V(p,q,\phi_h)\ . \end{equation}
This universality was understood before \cite{FKS,FK} as a consequence of
supersymmetry. Here the universal properties of the area of the horizon of the
extremal black holes are deduced only from the requirement of the regularity of
the configuration.

The equation of motion for $\phi^a$ is
\begin{eqnarray}
{D\phi^a \over D \tau^2} ={1\over 2}  {\partial V \over \partial \phi^a}
e^{2U} \ .
\end{eqnarray}
Near the horizon, taking into account that ${d \phi ^a \over d\omega } =\tau
{d \phi ^a \over d\tau  } =0$, we get
\begin{eqnarray}
{d^2 \phi^a \over d \tau^2} \rightarrow {1\over 2}  {\partial V \over \partial
\phi^a}
\left( {4\pi \over  \tau^2A} \right)  \ .
\end{eqnarray}
The solution of this equation near the horizon is
\begin{equation}
\phi^a = \left( {2 \pi \over A} \right) {\partial V \over \partial \phi^a} \log
\tau
+ \phi^a_h
\label{critical}\end{equation}
Above we have omitted a term linear in $\tau$ since it will lead to the
singular dilaton at the horizon.
Equation (\ref{critical}) shows that unless the derivative of the scalar
potential over the scalar field vanishes near the horizon, one can not have a
regular value of the scalars near the horizon. Thus
\begin{equation}
 \left({\partial V \over \partial \phi^a}\right)_h=0 \ .
\end{equation}
Finally we note that the behavior at infinity, at $\tau \rightarrow 0$, is
$U\rightarrow M\tau$ leads to the following constraint between the black hole
mass, scalar charges, scalar metric and the potential, valid regardless of
whether the hole is extreme or not,
\begin{equation}
M^2 + G_{ab} \Sigma ^a \Sigma ^b - V(p,q, \phi^a_{\infty} ) =  c^2 \ ,
\end{equation}
where  $\phi^a_\infty$ are the values of the scalars at spatial
infinity and   the scalar charges are defined via the expansion of the scalars
at
infinity.   Of course in the extreme limit we set $c=0$.
The double-extreme black holes have a vanishing scalar charge and the constant
fixed value of scalars everywhere:
\begin{eqnarray}
M^2 &=& V(p,q, \phi^a_{fix}) \\
{A\over 4\pi} &=& V(p,q, \phi^a_{fix})
\end{eqnarray}
and the fixed value of the scalars is defined by the extremization of the
potential
\begin{equation}
\left( {\partial V ( \phi,  p,q) \over \partial \phi^a }\right)_{fix}= 0
\end{equation}
Note that in this description of the extremal and double-extremal black holes
there was no use of supergravity and/or supersymmetry. We have used  the
bosonic field equations of the theory and the requirement that the extremal
configuration is regular near the horizon, including the regularity of scalars.
In the next section we will specify this study to the case of N=2 supergravity
and special geometry.

We have arrived at the following picture. We may associate with each critical
point $\phi ^a_{ fix}$ of the potential $V(\phi, p,q)$ on the manifold of
scalars ${\cal M}_\phi$
a supersymmetric Bertotti-Robinson vacuum state. Extreme black hole solutions
correspond to dynamical trajectories  in the moduli  space ${\cal M}_\phi$
starting from the  point $\phi ^a _\infty$ at spatial infinity and ending
on a critical point $\phi ^a_{fix}$. Double extreme holes
with frozen moduli
correspond to trivial point trajectories. One could  also consider trajectories
running
between two different critical points but these would not correspond to
asymptotically
flat solutions. Thus the extreme solutions may be said
to spatially interpolate between different vacua. Finding explicit trajectories
which effect this interpolation and which satisfy the second order dynamical
equations of motion
is in general quite difficult, although a number of solutions are known.
However   we shall see shortly that using
special  geometry  one is able  to reduce this problem to the easier one of
finding the solutions of a set of first order differential
describing the steepest descent curves of another potential function
whose physical significance is that it determines the central charge.

\section{Critical Points of the Central Charge and of the Potential in Special
Geometry}

Now we consider the special case for which the scalars field manifold ${\cal
M}_4$ is a Kahler manifold with complex coordinates $z^i$ and Kahler potential
$K$ so that
\begin{equation}
G_{ab} d \phi^a d \phi^b = {\partial ^2 K \over \partial z^i \partial \bar z^n
} dz^i d \bar z^{n}\ .
\end{equation}
The bosonic part of the action of N=2 supergravity interacting with some number
of vector multiplets is \footnote{Here we follow notation of \cite{KSW} where
the  black holes  were studied in the context of the special geometry. In this
context, as different from eq. (\ref{scalaraction}) the vector field strength
has additional factor 1/2.}
\begin{equation}  -{R\over 2} + G_{i\bar j} \partial_ \mu z^i   \partial_\nu
\bar z^{\bar j}
g^{\mu\nu} +
 {\rm Im} {\cal N}_{\Lambda \Sigma} {\cal F}^{\Lambda}_{\mu \nu}  {\cal
F}^{ \Sigma}_{\lambda \rho}  g^{\mu \lambda} g^{\nu \rho}
+ {\rm Re} {\cal N}_{\Lambda \Sigma} {\cal F}^{\Lambda}_{\mu \nu}{}^*{\cal
F}^{ \Sigma}_{\lambda \rho} g^{\mu \lambda} g^{\nu \rho}
 \ ,
\label{scalaraction2}\end{equation}
Here the positive definite metric $G_{i\bar j}$ on the scalar manifold as well
as scalar dependent negative definite vector couplings  ${\rm Re} {\cal
N}_{\Lambda \Sigma}$ and  ${\rm Re} {\cal N}_{\Lambda \Sigma}$
can be derived from the prepotential or from the symplectic section which
defines a particular N=2 theory.
The symplectic invariant $I_1$ of the special geometry
 constructed in \cite{Cer1}
\begin{equation}
I_1 = |Z(z,p,q)|^2 + |D_i Z(z,p,q)|^2=- {1\over 2}(p,q)   \left |\matrix{
 {\rm Im} {\cal N} + {\rm Re} {\cal N}  {\rm Im} {\cal N} ^{-1} {\rm Re} {\cal
N}  &-{\rm Re} {\cal N}  {\rm Im} {\cal N}^{-1}  \cr-
 {\rm Im} {\cal N}^{-1} {\rm Re} {\cal N} &  {\rm Im} {\cal N}^{-1} \cr
}\right |
  \left (\matrix{
p\cr
q\cr
}\right ),
\end{equation}
 can be identified with the potential
\begin{equation}
V(p,q, \phi^a ) = I_1 = |Z(z,p,q)|^2 + |D_i Z(z,p,q)|^2 \ .
\end{equation}
{}From eqs. (\ref{pot}), (\ref{Matrix}) of the present paper and from eq. (56)
of \cite{Cer1} it  follows that the identification requires that
\begin{equation}
\nu + i \mu = - {\cal N }= -  {\rm Re} {\cal N } - i  {\rm Im} {\cal N }
\end{equation}
Here $Z$ is the central charge \cite{Cer}, the charge of the graviphoton in N=2
supergravity and $D_i Z$ is the Kahler covariant derivative of the central
charge:
\begin{equation}
Z(z, \bar z, q,p) = e^{K(z, \bar z)\over 2}
(X^\Lambda(z)  q_\Lambda - F_\Lambda(z) \, p^\Lambda)= (L^\Lambda
q_\Lambda -
M_\Lambda p^\Lambda) \ .
\label{central}\end{equation}

We will use the abbreviation
\begin{equation}
\left| {d z \over d \tau} \right|^2 = G_{ab} {d \phi^a \over d \tau} {d \phi^b
\over d\tau} = {\partial ^2 K \over \partial z^i \partial \bar z^n } {dz^i
\over d \tau} {d \bar z^{n} \over d \tau}\ .
\end{equation}

The first symplectic invariant of special geometry $I_1$ is positive definite.
Our one-dimensional Lagrangian is
\begin{equation}
 {\cal L} \left (U(\tau) , z^i(\tau) , \bar z^i(\tau) \right )= \left ({d U
\over d\tau}\right )^2 +   \left |{d z \over d\tau } \right |^2
 + e^{2U} \left( |Z(z,p,q)|^2 + |D_i Z(z,p,q)|^2 \right)
\ .
\label{lagr}\end{equation}
The constraint becomes
\begin{equation}
  \left ({d U
\over d\tau}\right )^2 +   \left |{d z \over d\tau } \right |^2
 - e^{2U} \left( |Z(z,p,q)|^2 + |D_i Z(z,p,q)|^2 \right) =c^2
\ .
\label{constr1}\end{equation}
At infinity, at $\tau \rightarrow 0$,
$U\rightarrow M\tau$ we get
\begin{equation}
M ^2 (z_{\infty}, \bar z_{\infty} , p,q )
 -   |Z(z_{\infty} ,p,q)|^2  =c^2
\ .
\end{equation}
The BPS configuration has the mass equal to the central charge in
supersymmetric theories
\begin{equation}
M ^2 (z_{\infty}, \bar z_{\infty} , p,q )
 -   |Z(z_{\infty} ,p,q)|^2   \ ,   \qquad  c=0
\
\label{BPS}\end{equation}
since the second and the fourth term in the l.h.s. of eq. (\ref{constr1})
cancel at $\tau \rightarrow 0$ and $U\rightarrow M\tau$.

By using some properties of special geometry and of the central charge and its
covariant derivatives one can rewrite the action as
\begin{equation}
 {\cal L}= \left ({d U \over d\tau} \pm e^U |Z| \right )^2 +   \left |{d z^i
\over d\tau } \pm e^U G^{i\bar k} \bar D_{\bar k} \bar  Z \right |^2 \pm
 2 {d  \over d\tau } \left ( e^U |Z|  \right )  \ .
\label{lagr1}\end{equation}

Thus we may solve the second order equations by making the ansatz that the
following first order equations  hold \footnote{These first order equations
were derived
using supersymmetry in \cite{FKS,fre}. The derivation here is new and does not
use supersymmetry.}
\begin{eqnarray}
{d U \over d\tau}&=&e^U |Z| \\
{d z^i  \over d\tau } &=&e^U G^{i\bar k} \bar D_{\bar k} |Z|
\end{eqnarray}
These first order equations immediately give (by evaluating at infinity)
\begin{eqnarray}
M&=& |Z| (z_{0}, p,q)\ , \\
\Sigma^i  &=& G^{i\bar k} \bar D_{\bar k} \bar  Z \ .
\end{eqnarray}
{}From the behavior at the horizon $\tau \rightarrow \infty$ we get
\begin{eqnarray}
\left( {A\over 4\pi}\right)^{1/2} &=& |Z| (z_{h}) \ , \\
 D^i Z(z_{h}) &=& 0 \ .
\end{eqnarray}
where also
\begin{equation}
\partial_i V = \partial_i \left( |Z(z)|^2 + |D_i Z(z)|^2\right)
\end{equation}
Thus we have recovered and specified in this framework of special geometry the
equations obtained earlier.
To study the critical points of the potential we need few identities from
special geometry \cite{special}-\cite{Cer1}.
\begin{eqnarray}\label{1}
D_i D_j Z &=& i C_{ijk} G^{k\bar k} \bar D_{\bar k} \bar Z \ , \\
\nonumber\\\label{2}
 D_i \bar D_{\bar j} \bar  Z &=&  G^{i\bar j}  \bar Z \ , \\
\nonumber\\\label{3}
\bar D_{\bar m} Z &=& 0 \ .
\end{eqnarray}
Here $C_{ijk} $ is a completely symmetric covariantly holomorphic tensor.

The critical points of the potential coincide with the critical point of the
central charge. Using the identities above one gets
\begin{equation}
\partial_i V = 2  (D_i  Z) \bar Z  + i C_{ijk}  G^{j\bar m}  G^{k \bar k} \bar
D_{\bar m} \bar Z \bar D_{\bar k} \bar Z \ .
\end{equation}
Thus indeed
\begin{equation}
 D_i  Z = \bar D_{\bar k} \bar Z =0 \qquad \Longrightarrow \qquad \partial_i V
= \bar \partial_{\bar i}  V=0 \ .
\end{equation}

In what follows we will address the problem: when is the extremum of the
central charge and of the potential  a minimum and when it is a maximum.
Note that the second covariant derivative of the moduli of the central charge
at the critical point coincides with the partial  (non-covariant) second
derivative.
We have to calculate
\begin{equation}
D_i D_j |Z|= \partial_i \partial_j |Z|  \qquad \bar D_{\bar i} D_j |Z|= \bar
\partial_{\bar i }\partial_j |Z| \qquad {\rm at} \qquad D_i  |Z|=  \partial_i
|Z|=0
\end{equation}
and the conjugate to this. We start with the second derivative of the moduli of
the central charge and use some identities of special geometry.
\begin{equation}
D_i D_j |Z|=  -{1\over 4} |Z|^{-3} (\bar Z)^2 D_i  Z D_j  Z + {1\over 2}
|Z|^{-1}
i C_{ijk} G^{k\bar k} \bar D_{\bar k} \bar Z \ .
\end{equation}

It follows that at the critical point
\begin{equation}
\partial_i \partial_j |Z| =  \bar \partial_{\bar i} \bar  \partial_{\bar j}
|Z| =0 \ .
\end{equation}
The mixed derivatives are
\begin{equation}
 \bar D_{\bar i} D_j |Z|= {1\over 4} |Z|^{-1}  \bar D_{\bar i} \bar  Z D_j  Z +
{1\over 2} G_{\bar i j} |Z| \ .
\end{equation}
It follows that at the critical point we get
\begin{equation}
\left ( \bar  \partial_i \partial_j |Z|\right) _{\rm cr}=  {1\over 2} G_{\bar i
j} |Z|_{\rm cr} \ .
\end{equation}

This means that whenever the metric is positive in the moduli space at the
critical point, the BPS mass reaches its minimum. However if the scalar metric
is singular and  changes the sign the connection established above only
signals that the BPS mass at the critical point reaches the maximum outside the
range of validity of  regular special geometry. In general when the metric
changes the sign we have some sort of a phase transition and there is a
breakdown of the
effective Lagrangian unless new massless states appear.

Let us now proceed with the evaluation of the second derivative of the
potential $V$ at the critical point where the value of the potential is given
by the square of the central charge at the critical point of the central
charge.
\begin{equation}
\bar \partial_{\bar i}  V= \partial_{ i}  V=0  \qquad V_{\rm cr} = |Z_{\rm
cr}|^2 \ .
\end{equation}
Upon some extensive use of the identities of special geometry we conclude that
at the critical points of the potential which are simultaneously the critical
points of the central charge  we get
 \begin{equation}
D_i D_j V=  0  \qquad {\rm at} \qquad D_i  Z=  \partial_i |Z|=0
\end{equation}
and
\begin{equation}
\left(  \bar D_{\bar i} D_j V\right)_{\rm cr}=  \left( \bar \partial_{\bar i}
\partial_j  V \right)_{\rm cr} = 2 G_{\bar i j} V_{\rm cr} \ .
\end{equation}
Thus again the sign of the second derivative of the potential is defined by the
sign of the metric on the moduli space at the critical point where the central
charge and the potential have vanishing derivatives.

Now we recall that the extremal condition for the central charge was
brought to an equivalent form of stabilization equation in \cite{FK,Dieter}
under
condition that the special geometry is not singular. We found that all moduli
from the vector multiplets become functions of electric and magnetic charges.
The stabilization equation of special geometry is \cite{FK,Dieter}
 \begin{equation}
\left (\matrix{
p^\Lambda\cr
q_\Lambda\cr
}\right )={ \rm Re} \left (\matrix{
2i \bar Z L^\Lambda\cr
2i \bar Z M_\Lambda\cr
}\right )  \equiv  { \rm Re} \pmatrix{2 i Y^I\cr  2 i F_J(Y)\cr}\ .
\label{stab}
\end{equation}
The fixed values of moduli of special geometry in terms of electric and
magnetic charges has been found in many examples before
\cite{FKS,FK,KSW,Dieter}. They
correspond to the fixed values to which scalar fields are attracted near the
black hole horizon. All known examples  solve both the extremality condition of
the central charge $\partial_i |Z|=0$ and the stabilization equation
(\ref{stab}). In fact they have been found mostly by solving the easier
equation: the stabilization equation (\ref{stab}):
\begin{equation}
(z^i)_{\rm cr} = z^i (p,q) \ .
\end{equation}
For many  of these solutions the entropy of N=2 black holes has been understood
from the microscopic point of view \cite{KLMS} via the counting of the string
or M-theory states.

The issue of the uniqueness of the critical points has not been studied before,
we have mainly focussed
on the goal of finding  critical points. Now that we have studied the
second derivatives of the central charge and of the potential at the critical
points one can address the issue of the uniqueness of the critical points
\footnote{The detailed analysis of the uniqueness issue of the critical points
(with particular examples) has been performed in the context of the
5-dimensional very special geometry  in \cite{new}. Here we outline how
this analysis may be extended to  4-dimensional special geometry.}. Under the
assumption that we
limit ourself exclusively to study only the
potentials and their critical points in the range of applicability of special
geometry (the scalar metric is strictly positive) we may expect that the
minimum of the potential, which is also the minimum of the BPS mass, is unique.
Indeed we have proved that the second
derivative at the critical point is positive when the scalar metric is positive
and the critical point is a minimum. This concerns any critical point in this
class and
therefore it has to be unique at least for the continuous branch of the
potential. However if we were to  relax the condition that the scalar metric is
positive we might find some disconnected branches of the potential
exhibiting some maxima and various other  phenomena. We refer the reader to the
5d case \cite{new} where we present particular examples of such situations. In
the 4d context the analogous investigation would amount to studying all
possible
branches of the potentials for various interesting theories and calculating the
value of the scalar metric at the critical points where it will become a
particular function of charges, according to stabilization the equation
(\ref{stab}) (in case of negative scalar metric the use of stability equation
(\ref{stab})  may be questionable, rather one may  rely on the
direct solutions of
extremality condition $\partial_i |Z|=0$ for exhibiting the critical point).

\section{Critical Points of Moduli Space of Extended  Supergravities
}

Here we will use the recent geometric formulation of extended supergravities
in which the duality symmetries of the theory are manifest \cite{AAF}. Below we
will present the minimal amount of information on this which will allow us
to describe the critical points of these theories. The reader is referred for
the details of the new construction to \cite{AAF}.

All $d=4, N>2$ supergravities have scalar fields whose kinetic Lagrangian is
described by sigma models of the form ${G\over H}$. Here $G$ is a non compact
group acting as an isometry group on the scalar manifold while $H$ is the
isotropy group. For $N\leq 4$ a supergravity can be coupled to matter. For such
cases the isotropy group is given by a direct product $H= H_{\rm Aut}\otimes
H_{\rm matter} $ of the automorphism group of the supersymmetry algebra $H_{\rm
Aut}$ and the isotropy group of the matter multiplets $H_{\rm matter} $. In
pure supergravity, when matter is absent $H$ is just $H_{\rm Aut}$. The
supergravity theory is completely specified in terms of the geometry of the
coset space and in particular in terms of the coset representatives $L$. For
the case of our interest  which are D=4 $N>2$ theories the group $G$ has to be
embedded into $Sp(2n,{\bf R})$ group. The construction therefore presents
symplectic sections of a  $Sp(2n_v,{\bf R})$ bundle over ${G\over H}$, given by
$f=(f^\Lambda_{AB}, h=h_{\Lambda AB})$ and $(f^\Lambda_I, h_{\Lambda_I}$). Here
${AB}$ are indices \footnote{ Upper $SU(N)$ indices label objects in the
complex conjugate representation of  $SU(N)$}. in the antisymmetric
representation   of $H_{\rm Aut}= SU(N)\times U(1)$ and $I$ is the index of the
fundamental representation of $H_{\rm matter} $. The graviphoton central
charges $Z_{AB}, \bar Z^{AB}$ and the matter charges $Z_I,  \bar Z^I$ are
defined as  linear combination of  quantized electric and magnetic charges and
moduli as follows:
\begin{eqnarray}
Z_{AB} &=&f^\Lambda_{AB}q_\Lambda - h_{\Lambda AB} p^\Lambda \ ,\\
\nonumber\\
Z_{I} &=&f^\Lambda_{I}q_\Lambda - h_{\Lambda I} p^\Lambda \ .
\end{eqnarray}

 The crucial observation which  will make it possible to establish a complete
universality of critical phenomena in extended supergravities is the following.
The manifestly symplectic form of supergravity supplies a simple and completely
general expression for the the black hole potential $V$ presented in eq.
(\ref{pot})  of this paper: it is given in  eq. (3.66) of \cite{AAF} upon
identification between the scalar couplings in  (\ref{scalaraction}) and in
manifestly  symplectic form of extended supergravities in \cite{AAF}.
\begin{equation}
V= {1\over 2} Z_{AB} \bar Z^{AB}+Z^I \bar Z_I \ .
\label{AAF}\end{equation}
The differential relations among charges follow from their definition  with the
use of a vielbein $P$ of ${G\over H}$. The embedded vielbein of the coset
consists of blocks:
\begin{equation}
{\cal P}= \pmatrix{
P_{ABCD} & P_{ABJ} \cr
P_{J AB} & P_{IJ} \cr
} \ .
\end{equation}
The differential equations which we will use for the study of the critical
points are
\begin{eqnarray}
\nabla  Z_{AB} &=& {1\over 2} \bar Z^{CD} P_{ABCD} + \bar Z_I P^I_{AB} \ , \\
\nonumber\\
\nabla  Z^{I} &=& {1\over 2} \bar Z^{CD} P^I_{CD} + \bar Z_J P^{JI} \ .
\label{full}\end{eqnarray}

 Now we have sufficient amount of the information on the properties of central
charges and moduli space of $N\geq 2$ theories to study the critical points.
We will first focus on pure supergravity without matter and afterwards will
study supergravities coupled to matter.

1. Critical points in pure $N\geq 2$  supergravities with $Z_I \equiv 0$.

The potential and the differential relation simplify:
\begin{equation}
V= {1\over 2} Z_{AB} \bar Z^{AB} \ .
\end{equation}

\begin{equation}
\nabla  Z_{AB} = {1\over 2} Z^{CD} P_{ABCD} \ .
\label{dif}\end{equation}
Using eq. (\ref{dif}) to find the derivatives of the central charge we get
\begin{equation}
\partial_i V ={1\over 4} \bar Z^{CD} \bar Z^{AB} P_{ABCD,i}  + {1\over 4}
Z_{AB} Z_{CD}P^{ABCD}{}_{,i}  \ . \end{equation}

Since $P_{ABCD}$ is completely antisymmetric the solution to this equation
exists where only one eigenvalue of the central charge matrix is non-vanishing
and other eigenvalues vanish \cite{FK}:
\begin{equation}
\partial_i V =0 \qquad {\rm at}  \qquad Z^{12} \neq 0  \ , \quad  Z_{\rm
other} = 0\ .
\end{equation}
The non-vanishing eigenvalue of the central charge matrix is the BPS mass at
the critical point.
 For the second derivative we get
 \begin{equation}
D_j \partial _i V\mid_{\partial_i V =0} = \partial_j \partial _i V = {1\over 2}
Z_{LM}
P^{ABLM}{}_{,j} \bar Z^{CD} P_{ABCD,i} \ .
\end{equation}
At present it is not clear if one can   simplify this expression and bring it
to the form close to what has been found in N=2 case. However, we expect
further developments in this direction.

\noindent 2. Matter coupled supergravities $N\geq 2$ with $Z_I \neq 0$.

The potential (\ref{AAF}) and the  differential relation (\ref{full}) now have
a contribution from the matter charges.  The critical point is defined by
\begin{equation}
\partial_i V ={1\over 4} \bar Z^{CD} \bar Z^{AB} P_{ABCD,i}  + {1\over 4}
Z_{AB} Z_{CD}P^{ABCD}{}_{,i} + {1\over 2} \bar Z_I P^I_{AB,i} \bar Z^{AB} +
{1\over 2}  Z^I P_{I,i} ^{AB}  Z_{AB}=0 \ .
\label{extreme}\end{equation}
The configurations with
\begin{equation}
Z^{12} \neq 0  \ , \quad  Z_{\rm  other} = 0 \quad Z^I= \bar Z_I=0
\end{equation}
solves the extremization condition (\ref{extreme}) for the potential.
The evaluation of the second derivative of the potential at the critical point
proceeds with the use of differential relations (\ref{full}) :
 \begin{equation}
 (\partial_j \partial _i V )_{cr}=  Z_{CD}  \bar Z^{AB}
(  P^{CDPQ}{}_{,j}  P_{ABPQ,i} + {1\over 2} P^{CD}_{Ij} P^I_{ABi}) \ .
\end{equation}
Again, it remains to be seen if by  using the coset space geometry one can
simplify this to a form analogous to a simple result in N=2 theory.

\section{Geometry, Thermodynamics and Critical Points}

In section two we described a general formalism for constructing
four-dimensional spherically symmetric solutions. In some ways the most
symmetrical formulation is the pure geodesics formulation involving
the fields ${\hat \phi}$ comprising the scalars $\phi^a$, the potentials for
the vectors $\psi^A, \chi_A$ and the Newtonian
potential $U$ all on an equal footing. This naturally arises from dimensional
reduction, three dimensions which automatically places $U$ and other
Kaluza-Klein
scalars and vectors on the same footing. Physically it is more convenient
to eliminate the potentials $\psi^A$ and $\chi_A$ in favour of their conjugate
conserved charges. This gives rise to the potential $V(p,q, \phi,)$
 while we  have the scalars $\phi^a$ and the Newtonian potential $U$
remain on the same footing and are described by a simple dynamical model
involving motion in the $U,\phi$ space with a potential. As we have seen
the essential properties of the extreme black holes, such as the
area of the event horizon  are given by the values of the potential
$V(q,p,\phi)$ thought of as a function on the moduli space of
scalars ${\cal M}_\phi$ at its critical points $\phi^a_{fix}$.
  Now it becomes especially interesting to ask how the mass $M$ and area
$A$ depend on the moduli. In particular about their second covariant
derivatives.
At this point it is worth recalling some standard geometrical ideas in ordinary
thermodynamics, see  \cite{rup} for a review.

 Some time ago  Weinhold  suggested using as a metric the Hessian of
the energy $M$, considered as a function of the $n+1$ extensive variables
$N^\mu = (S, N^a)$,  where $S$ is the entropy and $N^a$, $a= 1,\dots n$ are
conserved numbers. Note in this formulation of an ordinary gas
the volume is included as one of the $N^a$'s.
In the case of black holes, the $N^a$'s include conserved charges,
angular momenta and also
(see \cite{GKK})
the values $\phi^\infty$ of the moduli at infinity. Thus
in conventional thermodynamics the Weinhold metric $W_{\mu \nu}$ is given by
\begin{equation}
W_{\mu \nu} = {\partial M \over \partial N^ \mu \partial N^\nu } \ .
\end{equation}

In conventional thermodynamics the  Weinhold metric is positive semi-definite
because of the
fact that the energy is least among equilibrium  configurations with given
entropy $S$, and total numbers $N^a$.
Because
\begin{equation}
dM= TdS + \mu_a dN^a,
\end{equation}
the dual function to $M$ is of course $G$ the Gibbs free energy,
which should be thought of as function of the intensive variables
$\mu _\mu = (T, \mu _a) $
given by
\begin{equation}
G=M-TS-\mu _a N^a
\end{equation}
whence
\begin{equation}
dG = -SdT -N^a d\mu _a \ .
\end{equation}
Thus the inverse metric $ W^ {\mu \nu}$ is given by
\begin{equation}
W^{\mu \nu} = - {\partial G \over \partial \mu _\mu \partial \mu_ \nu }\ .
\end{equation}

Note that the negative sign arises because of the conventional
choice of sign made in the thermodynamics literature when defining the Legendre
transformation.

 Sometime after Weinhold,  Ruppeiner focussed attention on the entropy  $S$
considered as a function of the extensive variables $M$ and $N^a$. It is
convenient to define extensive charges $Q^\mu= (M, N^a)$ and
conjugate variables $\beta _\mu= ( { 1\over T}, -{ \mu _a \over T})$.
Ruppeiner observed that fluctuations of the system are governed by
the Ruppeiner metric
\begin{equation}
S_{\mu \nu} = -{\partial S \over \partial Q^ \mu \partial Q^\nu }\ .
\end{equation}
The inverse metric is given by
\begin{equation}
S^{\mu \nu} = {\partial \Gamma  \over \partial \beta _\mu   \partial \beta _\nu
}\ .
\end{equation}
where $\Gamma$ is the Legendre transform of the entropy $S$, i.e
\begin{equation}
\Gamma = { G \over T} = -S + {M \over T} -N^a { \mu _a \over T}\ .
\end{equation}
More symmetrically
\begin{equation}
S+ \Gamma= \beta _\mu Q ^\mu \ .
\end{equation}

An interesting question to ask is  how these two metrics  are related.
The answer is, perhaps, surprisingly, that they are conformally related and the
conformal
factor is the temperature,
in other words
\begin{equation}
W_{\mu \nu} d N^\mu d N^\nu = T S_{\mu \nu } dQ^\mu dQ^\nu \ .
\end{equation}

To see this note that
\begin{equation}
W_{\mu \nu} d N^\mu d N^\nu =dT \otimes _s  dS + d\mu_a \otimes _s dN^a
\end{equation}
while
\begin{equation}
-S_{\mu \nu} d Q^\mu d Q^\nu =d {1\over T}  \otimes _s  dM + d {\mu_a \over T }
\otimes _s dQ^a \ .
\end{equation}

It is interesting to observe  that it is the conformal geometry which
is physically relevant. Thus ratios of specific heats should be conformal
invariants.

We shall now relate these geometric thermodynamic ideas  to the work of this
paper.
 The first thing to note is that  that in general,
for non-extreme holes these metrics will not be positive
definite because of the fact that non-extreme black holes
with  have negative specific heats.

The heat capacity $C= \left ({\partial  M \over \partial T}\right)_{p,q,\phi}$
is related to the second derivative of the mass over the entropy at fixed
values of charges. For  dilaton  non-extreme black holes the change in the sign
of the heat
capacity
has been studied in \cite{G,Pat,US}. It has been shown for generic $U(1)^2$
that in the process of the black hole evaporation the temperature increases,
reaches the maximum and rapidly drops to zero when the mass of the black hole
reaches the value of the central charge.
Specific heat blows up when the temperature reaches the maximum and changes the
sign,
see Figs. 4, 6, 7 of \cite{US}. The change of the sign of the heat capacity
happens at the non-vanishing temperature and means that the corresponding
component of the Weinhold metric undergoes the same type of changes.

The second important point is that,
as pointed out in \cite{GKK},
 when considering black holes with scalars we must augment
the usual extensive thermodynamic variables such as the area $A$
and the conserved charges
$(q,p)$ with the values of the moduli at infinity $\phi_\infty$.

This has the consequence that the thermodynamic configuration space
is no longer flat ${ \bf R}^k \equiv (A, q,p)$ ( where $k$ is one plus
two times the number of electric charges, but becomes its product
with the scalar moduli space ${ \bf R}^k \times {\cal M}_\phi$.

It was shown in \cite{GKK} that the  thermodynamic
variables conjugate to to the moduli $\phi^a_\infty$ , i.e. the analogue
of the chemical potentials $\mu_a$ are minus the scalar charges,
i.e. one has the relation
\begin{equation}
dM=TdS+\psi^A dq_A + \chi_A dp^A - \Sigma _a d \phi ^a _\infty \ .
\end{equation}

The fact that the scalar moduli space ${\cal M}_\phi$ is no
 longer  flat complicates, but of course does  not invalidate,
the usual thermodynamic formalism  involving Legendre transformations
(technically one should  speak of Legendre submanifolds etc.)
and in particular one is
allowed  to extend the definition of the Weinhold and
Ruppeiner metrics to the scalar moduli space ${\cal M}_\phi$
provided we replace the ordinary derivative by
the covariant derivative with respect to the metric $G_{ab}$
on the moduli space.

In the present paper we have been considering extreme
black holes for which the temperature $T=0$ and it is the Weinhold metric
$W_{ab}$
which seems to be the more appropriate one to consider.
This may be defined by
\begin{equation}
W_{ab}= \nabla _a \nabla _b M(p,q,\phi) \ .
\end{equation}

The Ruppeiner metric governs fluctuations and naively
diverges (see relevant equation above) if $T\rightarrow 0$. This is in
agreement with the arguments presented in   \cite{Pat,US}
that near extreme the thermodynamics breaks down.
However one might consider a renormalized  definition
of the the  Ruppeiner metric
\begin{equation}
S_{ab} = { 1\over 4} \nabla _a \nabla _b A(p,q, \phi) \ .
\end{equation}

Note that if the mass $M$ considered as a function
of the scalars is at  a critical point the first derivative vanishes
and the covariant derivative may be replaced by the partial derivative.
For a general thermodynamic substance or for a general black hole
 one expects to be able to say very little about
the Weinhold metric. In the case of extreme black holes it
is given by
\begin{equation}
W_{ab}= { 1\over 2 \sqrt V} \nabla _a \nabla _b V \ .
\end{equation}
As to the Ruppeiner metric, because the area $A$ of the
event horizon depends only on the values of the scalars on the horizon
and is independent of their values at infinity it follows that
\begin{equation}
S_{ab}=0 \ .
\end{equation}

By contrast for black hole arising form special geometry we are able to make
rather more precise statements about the Weinhold metric.
We find the remarkable result that the Weinhold metric is proportional
to the metric $G_{ab}$ on the moduli space.

\section{Conclusion}

In conclusion we have found the properties of the critical points of the BPS
mass in the range of applicability of the special geometry.  Supersymmetric
states in the spectrum of N=2 theory have the properties that their mass equals
the central charge  $M^{BPS} = |Z|$. The central charge $Z$ is defined in a
generic point of moduli space to be a particular function of moduli and
electric and magnetic charges and therefore the BPS mass in N=2 theory is given
by
\begin{equation}
M^{BPS} = M ( z, \bar z ,  p,q) \ .
\end{equation}
 When the derivative of the BPS mass over the moduli at fixed values of
electric and magnetic charges vanishes we call this a critical point of the
moduli space.

\begin{equation}
\left( {\partial\over \partial z^i}  M (  z, \bar z ,  p,q) \right) _{\rm cr}
=0
\qquad \Longrightarrow \qquad z_{\rm cr}  = z (p,q) \ .
\end{equation}
 The critical value of the BPS mass coincides with the value of the entropy of
the black hole with the corresponding charges: $\pi M(p,q)_{\rm cr} =
S(p,q)$ .

In this paper we have calculated the  second derivatives of the BPS mass   at
the critical point.
The result is simple and universal for all possible N=2 supergravities
interacting with arbitrary number of vector multiplets. It   shows that the
second derivative is proportional
to the metric in the moduli space and the critical value of the BPS mass.

\begin{equation}
\left( {\partial\over \partial z^i }  { \partial \over \partial \bar z^j }   M
( z, \bar z,  p,q) \right)_{\rm cr} = {1\over 2} G_{i\bar j }( z_{\rm cr}, \bar
z_{\rm cr} )  M (p,q)_{\rm cr} \ .
\end{equation}

Thus as long as the values  of the mass $M_{\rm cr}$ is positive and the scalar
metric on the moduli space $(G_{i\bar j})_{\rm cr} $ is positive definite at
the
critical point,  the BPS mass  reaches its unique  minimum at the critical
point. This fact was already applied to the study of the energy of the bound
states of branes \cite{K} and it was pointed out that the extremum of the
central charge describes the  bound states with minimal energy.

If however, any of these two positivity conditions are violated, the analysis
based on regular special geometry on N=2 supersymmetric theories does not
apply: one has to include the possibility of vanishing moduli and vanishing BPS
mass \cite{strom}  and of various singularities of special geometry,
in particular the change in the sign of the metric of the scalar manifold. This
will extend the study performed here to the  interesting cases relevant to
possible ``phase transitions" between different vacua in different theories as
suggested in \cite{witten}.  The examples of such behavior in the context of
the 5-dimensional Calabi-Yau type black holes will be presented in \cite{new}.

An interesting outcome of our analysis is the relation
of the metric on the moduli space $ G _{ab}$ with the thermodynamic
metric $W_{\mu \nu}$ introduced by Weinhold \cite{rup}. For
a general thermodynamic system it would seem to be very
difficult to say much about
the Weinhold metric. In the present case we are dealing, quite literally,
with special geometry and   in the extreme case
we have found that they are proportional. It would be of interest
to extend this analysis to the non-extreme case and this we plan to do
in the future. One motivation for studying the Weinhold metric is
that one might imagine that in a more exact quantum theory of gravity
in which spacetime geometry may not play the same pre-eminent role that
it does in classical and semi-classical general relativity
one will still be able to talk
about the thermodynamic properties of "black holes" but at a more abstract
level. One needs therefore some principle to determine
the thermodynamic surface giving the equation of state
of the system. The thermodynamic properties
are encoded in the the Weinhold
metric.  In theories based on an underlying geometric structure,
such as $N=2$ theories which are based on special geometry it is not
unreasonable to hope that the metric on moduli space and the
Weinhold metric continue to be closely related in the full quantum
regime.

\normalsize

\section*{Acknowledgements}
We would like to thank E. Cremmer and R. D'Auria for discussions on the
analysis covered
in sect. 5 and A. Chou,  J. Rahmfeld, S-J. Rey, M. Shmakova and
 W.K. Wong for the discussions of the non-uniquness of critical  points and
discontinuity of the potentials. The work of S. F. was supported
in part by DOE grant DE-FGO3-91ER40662
and by European Commission TMR programme ERBFMRX-CT96-0045 (INFN, Frascati).  
R.
K.  was supported by the NSF grant PHY-9219345.

\vfill
\newpage

\end{document}